\documentclass{revtex4}%
\usepackage{amssymb}
\usepackage{amsfonts}
\usepackage{amsmath}
\usepackage{graphicx}%
\begin{document}
\title{On the ro-vibrational energies for the lithium dimer; maximum-possible
rotational levels.}
\author{Omar Mustafa}
\email{omar.mustafa@emu.edu.tr}
\affiliation{Department of Physics, Eastern Mediterranean University, G. Magusa, north
Cyprus, Mersin 10 - Turkey,}
\affiliation{Tel.: +90 392 6301078; fax: +90 3692 365 1604.}

\begin{abstract}
The Deng-Fan potential is used to discuss the reliability of the improved
Greene-Aldrich approximation and the factorization recipe of Badawi et al.'s
\cite{17} for the central attractive/repulsive core $J\left(  J+1\right)
/2\mu r^{2}$. The factorization recipe is shown to be a more reliable
approximation and is used to obtain the ro-vibrational energies for the
$a^{3}\Sigma_{u}^{+}$ - $^{7}$Li$_{2}$ dimer. For each vibrational state only
a limited number of the rotational levels are found to be supported by the
$a^{3}\Sigma_{u}^{+}$ - $^{7}$Li$_{2}$ dimer.

\textbf{Keywords: }Ro-vibrational energies, Lithium dimer. Maximum-possible
rotational levels.

\end{abstract}
\maketitle

\section{Introduction}

The discovery of Bose-Einstein condensation in some ultracold spin-polarized
states of the alkali lithium dimer $^{7}$Li$_{2}$ \cite{1} has encouraged
intensive experimental as well as theoretical studies on this system
\cite{2,3,4,5,6,7}. Whilst the Bose-Einstein condensation formation in $^{7}%
$Li$_{2}$ dimer is found to depend on the interaction potential of the lowest
triplet excited $a^{3}\Sigma_{u}^{+}$ state, its stability is observed to be
sensitive to the binding energy of the least bound vibrational state (among
the 11 vibrational states supported by the $a^{3}\Sigma_{u}^{+}$ - $^{7}%
$Li$_{2}$ potential) \cite{2,3,4}. Spectral analysis were carried out to
determine vibrational and rotational constants and dissociation energy for
this dimer \cite{3,5,6}. The transition probabilities, moreover, depend on the
molecular rotational-vibrational (ro-vibrational, hereinafter) levels. A
general analytical closed-form solution for molecular ro-vibrational energies
(with sufficient reliable accuracy in a broad range of the rotational and
vibrational quantum numbers) would be of great interest in Physics and/or
Chemistry, for it would allow substantial simplifications of the derivation of
molecular transition probabilities \cite{8}. The ro-vibrational energy levels
of the $a^{3}\Sigma_{u}^{+}$ - $^{7}$Li$_{2}$ dimer represent the core of the
current work.

In the literature, an empirical\ diatomic molecular\ potential energy
function, $U\left(  r\right)  $, necessarily and desirably satisfies the
conditions (cf,. e.g., \cite{9,10})%
\begin{equation}
U\left(  \infty\right)  -U\left(  r_{e}\right)  =D_{e}\text{ },\text{ }\left.
\frac{dU\left(  r\right)  }{dr}\right\vert _{r=r_{e}}=0,\text{ and \ }\left.
\frac{d^{2}U\left(  r\right)  }{dr^{2}}\right\vert _{r=r_{e}}=K_{e}=\left(
2\pi c\right)  ^{2}\mu\omega_{e}^{2}.
\end{equation}
Where $D_{e}$ is the dissociation energy, $r_{e}$ is the equilibrium bond
length, $c$ is the speed of light, $\mu$ is the reduced mass, and $\omega_{e}$
is the equilibrium harmonic oscillator vibrational frequency. The introduction
of a fourth condition $U\left(  r_{e}\right)  =0$ would only shift the
potential by a constant at the equilibrium bond length, but never violates the
three conditions above \cite{11,12,13,14,15,16,17,18,19} For example, the
Schi\"{o}berg \cite{15}, and the improved (or the shifted by a constant)
Manning-Rosen potentials share the Deng-Fan \cite{14} diatomic molecular
potential form
\begin{equation}
U\left(  r\right)  =D_{e}\left[  1-\frac{e^{\alpha r_{e}}-1}{e^{\alpha r}%
-1}\right]  ^{2}.
\end{equation}
Here $\alpha$ denotes the range of the potential and is obtained using the
last condition in (1) to read%
\[
\alpha=\beta+\frac{1}{r_{e}}W(-r_{e}\beta\,e^{-r_{e}\beta});\text{ }%
\beta=\sqrt{\frac{K_{e}}{2D_{e}}},
\]
where $\beta$ is often called the Morse constant \cite{19} and $W\left(
z\right)  $ is the Lambert function. The vibrational spectra of such a model
is exactly solvable and a closed form solution exists in the literature (e.g.,
\cite{11,12}). The main challenge lies, however, in dealing with the central
attractive/repulsive core $J\left(  J+1\right)  /2\mu r^{2}$ of the radial
spherically symmetric Schr\"{o}dinger equation%
\begin{equation}
-\frac{\hbar^{2}}{2\mu}\frac{d^{2}u_{\nu,J}\left(  r\right)  }{dr^{2}}+\left[
\frac{J\left(  J+1\right)  \hbar^{2}}{2\mu r^{2}}+U\left(  r\right)  \right]
u_{\nu,J}\left(  r\right)  =E_{\nu,J}u_{\nu,J}\left(  r\right)  ,
\end{equation}
with $\nu$ denoting the vibrational and $J$ denoting the rotational quantum numbers.

In their attempt to obtain the ro-vibrational spectra for the $a^{3}\Sigma
_{u}^{+}$ - $^{7}$Li$_{2}$ dimer, Liu and coworkers \cite{11} have, very
recently, used an improved Manning-Rosen empirical potential energy model
(2).\ However, to deal with the central attractive/repulsive core $J\left(
J+1\right)  /2\mu r^{2}$ (i.e., the rotational-vibrational coupling) they have
used the improved Greene-Aldrich approximation \cite{16}
\begin{equation}
\frac{1}{r^{2}}\approx\alpha^{2}\left(  \frac{1}{12}+\frac{e^{\alpha r}%
}{\left(  e^{\alpha r}-1\right)  ^{2}}\right)  ,
\end{equation}
and reported the ro-vibrational energy spectra in a closed analytical form as%
\begin{equation}
E_{\nu,J}=D_{e}+\frac{J\left(  J+1\right)  \hbar^{2}\alpha^{2}}{24\mu}%
-\frac{\hbar^{2}\alpha^{2}}{2\mu}\left(  \frac{\frac{2\mu}{\hbar^{2}\alpha
^{2}}D_{e}\left(  e^{2\alpha r_{e}}-1\right)  }{\Lambda}-\frac{\Lambda}%
{4}\right)  ^{2},
\end{equation}
with%
\begin{equation}
\Lambda=2\nu+1+\sqrt{\left(  1+2J\right)  ^{2}+\frac{8\mu}{\hbar^{2}\alpha
^{2}}D_{e}\left(  e^{\alpha r_{e}}-1\right)  ^{2}}.
\end{equation}
but never subjected it to a quantitative brute force numerical test for any
$J\neq0$. At this very point, it is obvious that the asymptotic behavior of
$e^{\alpha r}/\left(  e^{\alpha r}-1\right)  ^{2}$ as $\alpha r\rightarrow0$
would manifest the necessary improvement of the Greene-Aldrich approximation%
\[
\frac{1}{r^{2}}\approx\alpha^{2}\left(  \frac{e^{\alpha r}}{\left(  e^{\alpha
r}-1\right)  ^{2}}\right)
\]
into the Taylor series expansion%
\[
\frac{e^{\alpha r}}{\left(  e^{\alpha r}-1\right)  ^{2}}\approx\left(
\frac{1}{\alpha^{2}r^{2}}-\frac{1}{12}+O\left(  \alpha^{2}r^{2}\right)
\right)  ,
\]
which in turn leads to (4). However, one may wonder as to whether such an
approximation is an adequate representation of the rotational-vibrational
coupling term. Strictly speaking, if such an approximation sacrifices the
accuracy for large rotational quantum number $J>0$ then one should look for an
alternative approach and hope for a better and more adequate representation.

In the current proposal, we suggest Badawi et al.'s \cite{17} factorization
recipe%
\begin{equation}
\frac{r_{e}^{2}}{r^{2}}=C_{0}+\frac{C_{1}}{e^{\alpha r}-1}+\frac{C_{2}%
}{\left(  e^{\alpha r}-1\right)  ^{2}},
\end{equation}
in section 2, as an alternative approach and report a closed form analytical
solution for the ro-vibrational energy levels. Although a variant of algebraic
approaches are available in the literature (cf., e.g., Infeld and Hull
\cite{20}, Wybourne \cite{21}, and Iachello and Levine \cite{22,23}), we
recollect (in the same section) the supersymmetric quantization recipe used by
Jia et al. \cite{18} to obtain the ro-vibrational energies. Moreover, we
subject (in section 3) both approaches (4) and (7) into a quantitative brute
force numerical test and compare their accuracy performance with those of Roy
\cite{19}, who have used a generalized pseudospectral ( GPS) method to
calculate the ro-vibrational energies for six diatomic molecules. We choose
the $O_{2}$( $X^{3}\Sigma_{g}^{-}$ ) molecule for the sake of comparison.
Owing to the fact that only 11 vibrational levels are\ supported by the
$a^{3}\Sigma_{u}^{+}$ - $^{7}$Li$_{2}$ potential (here the Deng-Fan \cite{14}
diatomic molecular potential\ (2) is just one of the available options and
need not be the best one for the $^{7}$Li$_{2}$ dimer), one would intuitively
expect similar trends for the rotational levels. We shall see that only a
limited number of the rotational levels are supported by the $a^{3}\Sigma
_{u}^{+}$ - $^{7}$Li$_{2} $ potential for each available vibrational state.
This is also discussed in section 3. To the best of our knowledge, such a
study is not reported elsewhere. Section 4 is devoted for our concluding remarks.

\section{Ro-vibrational energies and supersymmetric quantization recipe}

In this section we recollect Jia et al.'s \cite{18} work on the 6-parametric
exponential-type one-dimensional potential, where a closed form exact energy
eigenvalues are obtained. For the sake of our study here, we use a
4-parametric potential and cast the Deng-Fan \cite{14} diatomic molecular
potential (2) as
\begin{equation}
U\left(  r\right)  =P_{1}+\frac{P_{2}}{e^{\alpha r}-1}+\frac{P_{3}}{\left(
e^{\alpha r}-1\right)  ^{2}},
\end{equation}
where%
\begin{equation}
P_{1}=D_{e}\text{ ; \ }P_{2}=-2D_{e}\left(  e^{\alpha r_{e}}-1\right)  \text{
; \ }P_{3}=D_{e}\left(  e^{\alpha r_{e}}-1\right)  ^{2}.\text{ }%
\end{equation}
Incorporating (7) and (8) into (3) one would write the effective potential as%
\begin{equation}
U_{eff}\left(  r\right)  =\frac{J\left(  J+1\right)  \hbar^{2}}{2\mu r^{2}%
}+U\left(  r\right)  =\tilde{P}_{1}+\frac{\tilde{P}_{2}}{e^{\alpha r}-1}%
+\frac{\tilde{P}_{3}}{\left(  e^{\alpha r}-1\right)  ^{2}},
\end{equation}
with%
\begin{equation}
\tilde{P}_{1}=P_{1}+\gamma C_{0}\text{ ; \ }\tilde{P}_{2}=P_{2}+\gamma
C_{1}\text{ };\text{ \ }\tilde{P}_{3}=P_{3}+\gamma C_{2};\text{ }%
\gamma=\text{\ }\frac{J\left(  J+1\right)  \hbar^{2}}{2\mu r_{e}^{2}}.
\end{equation}
and the values of the $C_{i}^{\prime}s$ are obtained in the following manner.
Let $y=\alpha\left(  r-r_{e}\right)  $ then with $\alpha r=y+u$ and
$\,u=\alpha r_{e}$ one implies that
\begin{equation}
\frac{r_{e}^{2}}{r^{2}}=\frac{1}{\left(  y/u+1\right)  ^{2}}\text{ \ and
\ }\frac{r_{e}^{2}}{r^{2}}=C_{0}+\frac{C_{1}}{e^{y+u}-1}+\frac{C_{2}}{\left(
e^{y+u}-1\right)  ^{2}}\text{.}%
\end{equation}
Retaining the first three terms of the Taylor's expansion near the equilibrium
internuclear distance $y\rightarrow0$ (i.e., $r\rightarrow r_{e} $) of both
expressions in (12) and equating coefficients of same power of $y$ one obtains%
\begin{align}
C_{0}  &  =1-\left(  \frac{1-e^{-u}}{u}\right)  ^{2}\left[  \frac{4u}%
{1-e^{-u}}-\left(  3+u\right)  \right]  \medskip,\\
C_{1}  &  =2\left(  e^{u}-1\right)  \left[  3\left(  \frac{1-e^{-u}}%
{u}\right)  -\left(  3+u\right)  \left(  \frac{1-e^{-u}}{u}\right)
^{2}\right]  \medskip,\\
C_{2}  &  =\left(  \frac{e^{u}-1}{u}\right)  ^{2}\left(  1-e^{-u}\right)
^{2}\left[  \left(  3+u\right)  -\frac{2u}{1-e^{-u}}\right]  \medskip.
\end{align}
Which are in exact accord with those reported in equation (4) of \cite{17}.

Under such potential parametric settings, one would use the supersymmetric
quantum recipe used by Jia et al.\cite{18} and follow, step-by-step, their
procedure for our Schr\"{o}dinger equation in (3), along with the effective
potential in (10). Namely, one should set their $P_{4}=P_{5}=0$ and their
$P_{1}$, $P_{3}$, and $P_{2}$ are our current $\tilde{P}_{1}$, $\tilde{P}_{2}%
$, and $\tilde{P}_{3}$, respectively. Hereby, we only cast the necessary
formulae where our superpotential would read%
\begin{equation}
\tilde{W}\left(  r\right)  =-\frac{\hbar}{\sqrt{2\mu}}\left(  \tilde{Q}%
_{1}+\frac{\tilde{Q}_{2}}{e^{\alpha r}-1}\right)  ,
\end{equation}
where the one-dimensional ground-state like wave function is given by%
\begin{equation}
\psi\left(  r\right)  =N\exp\left(  -\frac{\sqrt{2\mu}}{\hbar}\int\tilde
{W}\left(  r\right)  dr\right)
\end{equation}
Which, when substituted in (3) along with (10), would result in%
\begin{equation}
\tilde{Q}_{2}^{2}-\alpha\,\tilde{Q}_{2}=\frac{2\mu}{\hbar^{2}}\tilde{P}%
_{3}\Longrightarrow\tilde{Q}_{2}=\frac{\alpha}{2}\left(  1+\sqrt{1+\frac{8\mu
}{\hbar^{2}\alpha^{2}}\tilde{P}_{3}}\right)
\end{equation}%
\begin{equation}
2\tilde{Q}_{1}\tilde{Q}_{2}-\alpha\tilde{Q}_{2}=\frac{2\mu}{\hbar^{2}}%
\tilde{P}_{2}\Longrightarrow\tilde{Q}_{1}=\frac{1}{2\tilde{Q}_{2}}\left[
\frac{2\mu}{\hbar^{2}}\left(  \tilde{P}_{2}-\tilde{P}_{3}\right)  +\tilde
{Q}_{2}^{2}\right]
\end{equation}
and%
\begin{equation}
\tilde{Q}_{1}^{2}=\frac{2\mu}{\hbar^{2}}\left(  \tilde{P}_{1}-E_{0}\right)
\Longrightarrow E_{0}=\tilde{P}_{1}-\frac{\hbar^{2}}{2\mu}\left(  \frac
{1}{2\tilde{Q}_{2}}\left[  \frac{2\mu}{\hbar^{2}}\left(  \tilde{P}_{2}%
-\tilde{P}_{3}\right)  +\tilde{Q}_{2}^{2}\right]  \right)  ^{2}.
\end{equation}
Under such settings, the wave function is%
\begin{equation}
\psi\left(  r\right)  =N\,e^{\tilde{Q}_{1}r}\left(  \frac{e^{\alpha r}%
-1}{e^{\alpha r}}\right)  ^{\tilde{Q}_{2}/\alpha}%
\end{equation}
and the corresponding eigenvalues are%
\begin{equation}
E_{\nu,J}=\tilde{P}_{1}-\frac{\hbar^{2}\alpha^{2}}{2\mu}\left[  \frac
{\frac{2\mu}{\hbar^{2}\alpha^{2}}\left(  \tilde{P}_{3}-\tilde{P}_{2}\right)
}{-1-2\nu-\sqrt{1+\frac{8\mu}{\hbar^{2}\alpha^{2}}\tilde{P}_{3}}}%
-\frac{-1-2\nu-\sqrt{1+\frac{8\mu}{\hbar^{2}\alpha^{2}}\tilde{P}_{3}}}%
{4}\right]  ^{2},
\end{equation}
where%
\[
\tilde{P}_{3}-\tilde{P}_{2}=De\left(  e^{2\alpha r_{e}}-1\right)
+\gamma\left(  C_{2}-C_{1}\right)
\]
and%
\[
\tilde{P}_{3}=D_{e}\left(  e^{\alpha r_{e}}-1\right)  ^{2}+\gamma C_{2}%
\]
Hereby, it should be obvious to notice that this result, in (22), is in exact
accord with that of Liu and coworkers \cite{11,12}, in (5) and (6), for $J=0$
and hence $\gamma=0$ (i.e., only for the vibrational levels). This is also
documented in tables 1-5.

\section{Results and Discussion}

In connection with the central attractive/repulsive core $J\left(  J+1\right)
/2\mu r^{2}$, we now subject the improved Greene-Aldrich approximation
\cite{16}, in (4), and the Badawi et al.'s \cite{17} factorization recipe, in
(7), to a quantitative brute force numerical test. Hereby, we use the $O_{2}$(
$X^{3}\Sigma_{g}^{-}$ ) diatomic spectroscopic molecular parameters
$D_{e}=42041cm^{-1}$, $\omega_{e}=1580.2cm^{-1}$, and $r_{e}=1.207\,\mathring
{A}$ used by Roy \cite{19} and report the results in table 1. In the same
table, we show the energy shifts $\Delta_{Liu}=E_{Liu}-E_{Roy}$ and
$\Delta_{our}=E_{our}-E_{Roy}$ for Liu et al's \cite{11} results, in (5), and
for our results, in (22), compared with those of Roy's \cite{19} (GPS),
respectively. It is obvious that whilst the ro-vibrational energies reported
by Liu \cite{11} dramatically shift from those of Roy \cite{19} ((i.e.,
$\Delta_{Liu}$ grows from $\sim11cm^{-1}$ for $\left(  \nu,J\right)  =\left(
0,0\right)  $ to $\sim177cm^{-1}$ for $\left(  \nu,J\right)  =\left(
0,20\right)  $ and from $\sim151cm^{-1}\,$for $\left(  \nu,J\right)  =\left(
5,10\right)  $ to $\sim289cm^{-1}$for $\left(  \nu,J\right)  =\left(
5,20\right)  $) as $J$ increases, our energies remain at an almost constant
shift from Roy's results (i.e., $\Delta_{our}\sim11\,cm^{-1}$ for $\left(
\nu,J\right)  =\left(  0,0\right)  $ to $\left(  \nu,J\right)  =\left(
0,20\right)  $ and $\Delta_{our}\sim105\,cm^{-1}$for $\left(  \nu,J\right)
=\left(  5,10\right)  $ to $\left(  \nu,J\right)  =\left(  5,20\right)  $).
This observation would in turn imply that the factorization recipe (7) of
Badawi et al. \cite{17} is more stable and more adequate than that of
Greene-Aldrich approximation \cite{16} used by Liu et al. \cite{11}. Of course
one should expect such energy shifts because of the different forms of the
interaction potentials used. Roy \cite{19} have used Tietz-Hua potential
whereas the Deng-Fan \cite{14} is used here and also used by Liu \cite{11}.
Nevertheless, the Deng-Fan potential is shown to be equivalent to the improved
Manning-Rosen potential \cite{12,13}.

We now safely proceed with our calculations for the ro-vibrational energies
using the \textit{"reliable"} factorization recipe (7) of Badawi et al.
\cite{17}. In tables 2,3, and 4 we report the ro-vibrational energy levels for
the $a^{3}\Sigma_{u}^{+}$ - $^{7}$Li$_{2}$ dimer. Here, we have used the
$a^{3}\Sigma_{u}^{+}$ - $^{7}$Li$_{2}$ dimer spectroscopic molecular
parameters $D_{e}=333.690cm^{-1}$, $\omega_{e}=65.130cm^{-1}$, and
$r_{e}=4.173\,\mathring{A}$ as used by Liu et al. \cite{11}. In table 3,
nevertheless, one observes that the energies are listed up to $J=10$ for
$\nu=7$. Similar limited numbers of the ro-vibrational energies are also
observed in table 4. The binding energy is known to satisfy the condition%
\begin{equation}
E_{binding}=E_{\nu,J}-D_{e}<0.
\end{equation}
Therefore, when the energies $E_{\nu,J}$ approach the dissociation energy
$D_{e}=333.690cm^{-1}$ they would in fact signal the very existence of a
maximum possible rotational quantum number, $J_{\max}$, associated with a
corresponding vibrational quantum number, $\nu$. \ That is, for each of the
only available 11 vibrational states there is a maximum possible number of
rotational levels for the $a^{3}\Sigma_{u}^{+}$ - $^{7}$Li$_{2}$ molecular dimer.

In table 5, we report the feasibly \textit{"maximum-possible"}
\ ro-vibrational energy levels. For example, we observe that for $\nu=0$ the
\textit{"maximum-possible"} \ ro-vibrational energy level is $J_{\max}=34$,
for $\nu=1$ is $J_{\max}=32$, for $\nu=2$ is $J_{\max}=29$, for $\nu=3$ is
$J_{\max}=26$, and so on $J_{\max}$ gradually decreases as $\nu$ grows up to
the 11th vibrational state where only one rotational level is obtained at
$J_{\max}=1$. Of course, in judging on the \textit{"maximum-possible"}
\ ro-vibrational energy level we have taken into account the stability
patterns of our energy shifts discussed above for the $O_{2}$( $X^{3}%
\Sigma_{g}^{-}$ ) diatomic molecule (in table 1) and projected such stability
patterns for the $a^{3}\Sigma_{u}^{+}$ - $^{7}$Li$_{2}$ dimer (by comparing
our results for $\nu=0$ with the RKR (Rydberg-Klein-Rees) ones reported in
\cite{11}). We may very clearly observe that the improved Greene-Aldrich
approximation \cite{16}, in (4), used by Liu et al. \cite{11} ceases to
satisfy condition (23) at lower values of the rotational quantum number $J$
(documented in tables 3,4, and 5). Indeed, the the Deng-Fan \cite{14}
potential (2) (used here) may not be the best interaction potential to
describe the $a^{3}\Sigma_{u}^{+}$ - $^{7}$Li$_{2}$ molecular dimer. It had,
nevertheless, shown intuitive consistency with the common sense contemplation
on that each of the only available 11 vibrational states may, very well,
accommodate a limited number of rotational states.

\section{Concluding remarks}

In this study, we have considered the Deng-Fan \cite{14} potential (2) and
discussed the reliability of two available approximations for the central
attractive/repulsive core $J\left(  J+1\right)  /2\mu r^{2}$ (i.e., the
improved Greene-Aldrich approximation \cite{16}, in (4), and the Badawi et
al.'s \cite{17} factorization recipe, in (7)). We have studied and analyzed
the numerical outcomes of both approximations for $J>0$ using the $O_{2}$(
$X^{3}\Sigma_{g}^{-}$ ) diatomic molecule and compared the results with those
of Roy \cite{19}. As long as the rotational quantum number $J>0$ (especially
for $J>>0$) is in point, such a comparison suggested that the factorization
recipe (7) of Badawi et al. \cite{17} is more reliable than that of
Greene-Aldrich approximation \cite{16} used by Liu et al. \cite{11}. The
stability and reliability of which is documented in the almost constant energy
shifts $\Delta_{our}$ (obtained and listed in table 1). That is, $\Delta
_{our}\sim11\,cm^{-1}$ for $\left(  \nu,J\right)  =\left(  0,0\right)  $ to
$\left(  \nu,J\right)  =\left(  0,20\right)  $ and $\Delta_{our}%
\sim105\,cm^{-1}$for $\left(  \nu,J\right)  =\left(  5,10\right)  $ to
$\left(  \nu,J\right)  =\left(  5,20\right)  $, whereas $\Delta_{Liu}$ grows
up from $\sim11cm^{-1}$ for $\left(  \nu,J\right)  =\left(  0,0\right)  $ to
$\sim177cm^{-1} $ for $\left(  \nu,J\right)  =\left(  0,20\right)  $ and from
$\sim151cm^{-1}\, $for $\left(  \nu,J\right)  =\left(  5,10\right)  $ to
$\sim289cm^{-1}$for $\left(  \nu,J\right)  =\left(  5,20\right)  $.

On the other hand, we have used the same potential to study the ro-vibrational
energies for the $a^{3}\Sigma_{u}^{+}$ - $^{7}$Li$_{2}$ molecular dimer. We
have shown that only a limited number of the rotational levels is supported by
the $a^{3}\Sigma_{u}^{+}$ - $^{7}$Li$_{2}$ dimer. That is, the $\nu=0$
vibrational state accommodates only $34$ rotational levels, $\nu=1$
accommodates $32$, $\nu=2$ accommodates $29$, $\nu=3$ accommodates $26$,
$\nu=4$ accommodates $23$, $\nu=5$ accommodates $20$, $\nu=6$ accommodates
$16$, $\nu=7$ accommodates $11$, $\nu=8$ accommodates $9$, $\nu=9$
accommodates $5$, and $\nu=10$ accommodates only one rotational level$.$ To
the best of our knowledge, this has never been reported elsewhere.

\newpage

\begin{center}%
\begin{tabular}
[b]{l}%
Table 1:\\
\multicolumn{1}{c}{%
\begin{tabular}
[c]{ccccccc}\hline
$\ \nu$\ \ \  & $\ J$ \ \ \  & Ref. \cite{11} \ \ \ \  & $\ $Eq.(22)
\ \ \ \ \ \  & GPS \cite{19}\ \ \  & $\ \Delta_{Liu}$ \ \ \ \ \ \  &
$\Delta_{our}$ \ \ \\\hline
0 & 0 & \multicolumn{1}{l}{786.380} & \multicolumn{1}{l}{786.380} &
\multicolumn{1}{l}{775.074} & \multicolumn{1}{l}{11.306} &
\multicolumn{1}{l}{11.306}\\
& 1 & \multicolumn{1}{l}{789.941} & \multicolumn{1}{l}{789.153} &
\multicolumn{1}{l}{777.848} & \multicolumn{1}{l}{11.562} &
\multicolumn{1}{l}{11.305}\\
& 2 & \multicolumn{1}{l}{797.066} & \multicolumn{1}{l}{794.699} &
\multicolumn{1}{l}{783.395} & \multicolumn{1}{l}{13.671} &
\multicolumn{1}{l}{11.304}\\
& 10 & \multicolumn{1}{l}{982.257} & \multicolumn{1}{l}{938.844} &
\multicolumn{1}{l}{927.562} & \multicolumn{1}{l}{54.695} &
\multicolumn{1}{l}{11.282}\\
& 15 & \multicolumn{1}{l}{1213.661} & \multicolumn{1}{l}{1118.890} &
\multicolumn{1}{l}{1107.634} & \multicolumn{1}{l}{106.027} &
\multicolumn{1}{l}{11.256}\\
& 20 & \multicolumn{1}{l}{1533.912} & \multicolumn{1}{l}{1367.943} &
\multicolumn{1}{l}{1356.714} & \multicolumn{1}{l}{177.198} &
\multicolumn{1}{l}{11.229}\\
5 & 10 & \multicolumn{1}{l}{8412.365} & \multicolumn{1}{l}{8366.286} &
\multicolumn{1}{l}{8261.257} & \multicolumn{1}{l}{151.108} &
\multicolumn{1}{l}{105.029}\\
& 15 & \multicolumn{1}{l}{8635.437} & \multicolumn{1}{l}{8534.849} &
\multicolumn{1}{l}{8429.967} & \multicolumn{1}{l}{205.470} &
\multicolumn{1}{l}{104.882}\\
& 20 & \multicolumn{1}{l}{8944.154} & \multicolumn{1}{l}{8768.000} &
\multicolumn{1}{l}{8663.303} & \multicolumn{1}{l}{280.851} &
\multicolumn{1}{l}{104.679}\\\hline
\end{tabular}
}%
\end{tabular}

\medskip%

\begin{tabular}
[b]{l}%
Table 2\\%
\begin{tabular}
[c]{cclll}\hline
$\ \nu$ \  & $\ J$ \  & Ref.\cite{11} \ \ \  & Eq.(22) \ \  & RKR
\cite{11}\ \\\hline
0 & 0 & 31.7694 & 31.7694 & 31.857\\
& 1 & 32.4900 & 32.3035 & \\
& 2 & 33.9311 & 33.3714 & \\
& 3 & 36.0919 & 34.9720 & \\
& 4 & 38.9717 & 37.1038 & \\
& 5 & 42.5692 & 39.7651 & \\
& 10 & 71.2648 & 60.9108 & \\
& 15 & 117.5735 & 94.7215 & \\\hline
1 & 0 & 90.3292 & 90.3292 & 90.453\\
& 1 & 91.0254 & 90.8283 & \\
& 2 & 92.4177 & 91.8260 & \\
& 3 & 94.5054 & 93.3215 & \\
& 4 & 97.2877 & 95.3132 & \\
& 5 & 100.7635 & 97.7991 & \\
& 10 & 128.4858 & 117.5415 & \\
& 15 & 173.2177 & 149.0671 & \\\hline
\end{tabular}%
\begin{tabular}
[c]{cclll}\hline
$\ \nu$ & $\ J$ \  & Ref.\cite{11} \ \ \ \  & Eq.(22) \ \ \ \ \  & RKR
\cite{11}\ \\\hline
2 & 0 & 142.3939 & 142.3939 & 142.523\\
& 1 & 143.0660 & 142.8583 & \\
& 2 & 144.4101 & 143.7867 & \\
& 3 & 146.4255 & 145.1782 & \\
& 4 & 149.1115 & 147.0311 & \\
& 5 & 152.4668 & 149.3436 & \\
& 10 & 179.2270 & 167.6973 & \\
& 15 & 222.4000 & 196.9618 & \\\hline
3 & 0 & 188.0375 & 188.0375 & 188.240\\
& 1 & 188.6858 & 188.4676 & \\
& 2 & 189.9822 & 189.3274 & \\
& 3 & 191.9262 & 190.6159 & \\
& 4 & 194.5168 & 192.3315 & \\
& 5 & 197.7531 & 194.4724 & \\
& 10 & 223.5620 & 211.4519 & \\
& 15 & 265.1939 & 238.4792 & \\\hline
\end{tabular}
\\\hline
\end{tabular}
\medskip\medskip%

\begin{tabular}
[b]{l}%
Table 3\\%
\begin{tabular}
[c]{cclll}\hline
$\ \nu$ & $\ J$ \  & Ref. \cite{11} \ \ \  & $\ $Eq.(22) \ \  & RKR
\cite{11}\ \ \\\hline
4 & 0 & 227.332 & 227.332 & 227.679\\
& 1 & 227.9573 & 227.7287 & \\
& 2 & 229.2066 & 228.5206 & \\
& 3 & 231.0799 & 229.7072 & \\
& 4 & 233.5763 & 231.2870 & \\
& 5 & 236.6949 & 233.2580 & \\
& 10 & 261.5633 & 248.8779 & \\
& 15 & 301.6712 & 273.6914 & \\\hline
5 & 0 & 260.350 & 260.350 & 260.837\\
& 1 & 260.9517 & 260.7128 & \\
& 2 & 262.1544 & 261.4374 & \\
& 3 & 263.9578 & 262.5233 & \\
& 4 & 266.3611 & 263.9687 & \\
& 5 & 269.3633 & 265.7717 & \\
& 10 & 293.3016 & 280.0462 & \\
& 15 & 331.9027 & 302.6694 & \\\hline
\end{tabular}%
\begin{tabular}
[c]{cclll}\hline
$\ \nu$ & $\ J$ \  & Ref. \cite{11} \ \ \  & Eq.(22) \ \ \ \  & RKR
\cite{11}\ \\\hline
6 & 0 & 287.160 & 287.160 & 287.665\\
& 1 & 287.7389 & 287.4898 & \\
& 2 & 288.8955 & 288.1480 & \\
& 3 & 290.6299 & 289.1341 & \\
& 4 & 292.9411 & 290.4465 & \\
& 5 & 295.8281 & 292.0833 & \\
& 10 & 318.8468 & 305.0266 & \\
& 15 & 355.9576 & 325.4829 & \\\hline
7 & 0 & 307.832 & 307.832 & 308.098\\
& 1 & 308.3877 & 308.1285 & \\
& 2 & 309.4988 & 308.7209 & \\
& 3 & 311.1647 & 309.6083 & \\
& 4 & 313.3849 & 310.7892 & \\
& 5 & 316.1581 & 312.2614 & \\
& 10 & 338.2671 & 323.8878 & \\
&  &  &  & \\\hline
\end{tabular}
\\\hline
\end{tabular}
\medskip\medskip%

\begin{tabular}
[b]{l}%
Table 4\\%
\begin{tabular}
[c]{cclll}\hline
$\ \nu$ & $\ J$ \  & Ref.\cite{11} \ \ \ \ \  & $\ $Eq.(22) \ \ \  & RKR
\cite{11}\ \\\hline
8 & 0 & 322.432 & 322.432 & 322.155\\
& 1 & 322.9654 & 322.6962 & \\
& 2 & 324.0314 & 323.2236 & \\
& 3 & 325.6298 & 324.0134 & \\
& 4 & 327.7599 & 325.0640 & \\
& 5 & 330.4205 & 326.3735 & \\
& 6 & 333.6105 & 327.9394 & \\
& 7 & 337.3283 & 329.7585 & \\
& 8 & 341.5722 & 331.8272 & \\\hline
9 & 0 & 331.027 & 331.027 & 330.170\\
& 1 & 331.5383 & 331.2592 & \\
& 2 & 332.5598 & 331.7222 & \\
& 3 & 334.0913 & 332.4154 & \\
& 4 & 336.1322 & 333.3373 & \\\hline
10 & 0 & 333.683 & 333.683 & 333.269\\\hline
\end{tabular}
\\\hline
\end{tabular}

\medskip%
\begin{tabular}
[b]{l}%
Table 5\\%
\begin{tabular}
[c]{ccll}\hline
$\ \nu$ \ \  & $\ J$ \ \  & Ref.\cite{11} \ \ \ \  & $\ $Eq.(22)
\ \ \ \ \ \ \ \ \\\hline
0 & 20 & 181.0396 & 140.3931\\
& 25 & 261.0387 & 196.7792\\
& 30 & 356.7841 & 262.3569\\
& 34 & 444.0906 & 320.1449\\
1 & 20 & 234.5100 & 191.5654\\
& 25 & 311.7476 & 243.8808\\
& 30 & 404.1552 & 304.4789\\
& 32 & 445.1631 & 330.6137\\
2 & 20 & 281.5434 & 236.3203\\
& 25 & 356.0508 & 284.6083\\
& 29 & 426.2105 & 328.6431\\
3 & 20 & 322.2125 & 274.7310\\
& 25 & 394.0204 & 319.0345\\
& 26 & 410.0931 & 328.7317\\\hline
\end{tabular}%
\begin{tabular}
[c]{ccll}\hline
$\ \nu$ \ \  & $\ J$ \ \  & Ref.\cite{11} \ \ \ \  & $\ $Eq.(22)
\ \ \ \ \ \ \ \ \\\hline
4 & 20 & 356.5889 & 306.8700\\
& 23 & 396.4097 & 330.3157\\
5 & 20 & 384.7428 & 332.8067\\
6 & 16 & 365.0368 & 330.4078\\
7 & 11 & 344.3232 & 327.0375\\
8 & 9 & 346.3401 & 334.1416\\
9 & 5 & 338.6815 & 334.4858\\
10 & 1 & 334.1715 & 333.8826\\
&  &  & \\
&  &  & \\
&  &  & \\
&  &  & \\
&  &  & \\
&  &  & \\\hline
\end{tabular}
\\\hline
\end{tabular}

\end{center}

\textbf{Tables captions:}

\textbf{Table 1:} Ro-vibrational energies $E_{\nu,J}$ (in $cm^{-1}$ units) for
$O_{2}$( $X^{3}\Sigma_{g}^{-}$ ), where $\Delta_{Liu}$and $\Delta_{our}$
denote the energy shifts for Liu et al's \cite{11} in Eq.(5) and our results
from Eq.(22) compared with those of Roy's \cite{19} (GPS), respectively.

\textbf{Table 2: }Ro-vibrational energies $E_{\nu,J}$ (in $cm^{-1}$ units) for
$^{7}$Li$_{2}$( $a^{3}\Sigma_{u}^{+}$ ) with $\nu=0,1,2,$ and $3$. Our results
from Eq.(22) are compared with those of Liu et al's \cite{11} in Eq.(5) and
those of RKR \cite{11} whenever possible.

\textbf{Table 3: }Ro-vibrational energies $E_{\nu,J}$ (in $cm^{-1}$ units) for
$^{7}$Li$_{2}$( $a^{3}\Sigma_{u}^{+}$ ) with $\nu=4,5,6,$ and $7$. Our results
from Eq.(22) are compared with those of Liu et al's \cite{11} in Eq.(5) and
those of RKR \cite{11} whenever possible.

\textbf{Table 4: }Ro-vibrational energies $E_{\nu,J}$ (in $cm^{-1}$ units) for
$^{7}$Li$_{2}$( $a^{3}\Sigma_{u}^{+}$ ) with $\nu=8,9$ and $10$. Our results
from Eq.(22) are compared with those of Liu et al's \cite{11} in Eq.(5) and
those of RKR \cite{11} whenever possible.

\textbf{Table 5: }Ro-vibrational energies $E_{\nu,J}$ (in $cm^{-1}$ units) for
$^{7}$Li$_{2}$( $a^{3}\Sigma_{u}^{+}$ ). Our results from Eq.(22) are compared
with those of Liu et al's \cite{11} in Eq.(5). For each value of $\nu$ we show
the \textit{"maximum-possible"} \ rotational quantum number $J$ (i.e.,
$J_{\max}$ is the last value of $J$ for each $\nu$).

\end{document}